\documentclass[a4paper,oneside,11pt]{article}
\usepackage{epsfig,multicol}
\usepackage{graphicx}
\usepackage{amsmath,amssymb}
\usepackage{mathrsfs}

\begin{document}
\setcounter{page}{1}

\begin{center}
{\Large \bf Addendum: Nonlinear integral equations for the sausage model
(2017 J.Phys. A50 314005)}
\end{center}
\vskip .8cm \centerline{\bf Changrim  Ahn$^a$, Janos Balog$^b$, and Francesco Ravanini$^{c,d}$ }

\vskip 10mm

\centerline{\sl $^a$Department of Physics} \centerline{\sl Ewha Womans University}
 \centerline{\sl DaeHyun 11-1, Seoul 120-750, S. Korea}

\vskip 0.8cm

\centerline{\sl $^b$Institute for Particle and Nuclear Physics}
\centerline{\sl Wigner Research Centre for Physics, MTA Lend\"ulet Holographic QFT Group} 
\centerline{\sl 1525 Budapest 114, P.O.B. 49, Hungary}

\vskip 0.8cm

\centerline{\sl $^c$Department of Physics and Astronomy}
\centerline{\sl University of Bologna}
\centerline{\sl Via Irnerio 46, 40126 Bologna, Italy}
\vskip 0.3cm

\centerline{\sl $^d$ Istituto Nazionale di Fisica Nucleare, Sezione di Bologna}
\centerline{\sl Via Irnerio 46, 40126 Bologna, Italy}

\vskip 9.6ex

\begin{center}
{\bf Abstract}
\end{center}
We complete the derivation of the sausage model NLIE by giving a proof of the crucial relation (3.24) of the original paper based on the analytic properties of $Q$ and $\bar Q$.

\vskip .5cm
\section{Introduction}

In ref. \cite{us}, here below referred as I, we have written the set of
Non-Linear Integral Equations (NLIEs) governing the finite size effects of the
vacuum as well as the thermodynamics for the integrable deformation of
$O(3)$ non-linear sigma model (NLSM), getting it from a manipulation,
inspired by those introduced years ago by J. Suzuki \cite{Junji1,Junji2},
of the larger set of Thermodynamic Bethe Ansatz (TBA) equations of the model,
known since the original paper by Fateev, Onofri and Zamolodchikov \cite{FOZ}. 
However, one can realize that (I3.24)\footnote{Here we refer to the
equations of I as (Ix.xx), for example eq. (3.24) of I is referred as
(I3.24). Definitions, notation and symbols are as defined in I.},
a crucial relation in our derivation of the sausage model NLIE, is
not well-defined because neither $Q$ nor $\bar Q$ are analytic on the
real axis. Hence $\tilde Q$ and $\tilde{\bar Q}$ cannot be interpreted as
Fourier transforms along the real line\footnote{We thank Prof. J. Suzuki
for pointing this out.}.

In this Addendum we examine this problem carefully and show that the derivation of the sausage model NLIE remains valid in spite of this potential difficulty .

\section{Analyticity strips}

Our starting point is that the sausage model $Y$-system for the ground state has constant
solution in the infinite volume limit
$\ell=mr\to\infty$ :
\begin{equation}
y_k=k(k+2),\quad k=1,\dots,N-2;\qquad y_N=y_{N-1}=N-1; \qquad y_0=0.
\end{equation}
The corresponding $T$-system solution is
\begin{equation}
T_k=k+1,\quad k=1,\dots,N-1
\end{equation}
and
\begin{equation}
A=\bar A=2.
\end{equation}
For (I3.13-14) we choose the bounded solutions
\begin{equation}
Q=\bar Q=1.
\end{equation}
The other linearly independent solutions of the second order difference
equations (I3.13) and (I3.14) are $Q=\bar Q=\theta$, but these are not
bounded.

We assume that we have solved the TBA equations for finite (but large) volume 
\begin{equation}
y_a(\theta)=\exp\left\{\sum_b\frac{I_{ab}}{2\pi}\int_{-\infty}^\infty
\frac{{\rm d}\theta^\prime}{\cosh(\theta-\theta^\prime)}
L_b(\theta^\prime)\right\},
\quad a=1,\dots,N; \quad y_0={\rm e}^{-\ell\cosh\theta}y_1(\theta),
\end{equation}
where $I_{ab}$ is the incidence matrix of the sausage model TBA diagram 
(including the massive node) and $L_a=\log Y_a$. All $y_a$ functions are 
defined originally along the real line, where they are real and positive.

The shifts of the left-hand side of the Y-system equations (I3.1-3) along
${\rm Im}\,\theta$, often referred to as TBA steps, are $\pm i\pi / 2$, 
so it is convenient to use the notation $(\alpha,\beta)$
indicating the strip
\begin{equation}
\frac{\pi}{2}\alpha<{\rm Im}\,\theta<\frac{\pi}{2}\beta.
\end{equation}
The above TBA equations themselves allow us to analytically continue the 
$Y$-functions to the strip $(-1,1)$ and we can see that 
all $y_a$ functions ($a=1,\dots, N$) are analytic and non-zero (ANZ) in this strip 
for large volume and they must be 
close to the constant solution.
$y_0$ is also ANZ in this strip and it is
uniformly small in the strip $(-1+\epsilon,1-\epsilon)$, where $\epsilon$ is some fixed, 
small, but not infinitesimal number. 
We will abbreviate this property by ANZC, 
meaning that it is ANZ and close to a constant solution. Then,
\begin{equation}
y_a {\rm \ is \ ANZC\  in\ } (-1,1) \quad{\rm for}\quad a=1,\dots,N;\qquad
y_0 {\rm \ is \ ANZC\  in\ } (-1+\epsilon,1-\epsilon).
\end{equation}
We can further extend these \lq\lq good" strips
for the $Y$-functions and also for the corresponding $T$-system
using the $Y$-system equations. In the appendix we show that
\begin{equation}
T_k {\rm \ is \ ANZC\  in\ } (-k-1+\epsilon,k+1-\epsilon),\qquad 
k=1,\dots,N-1.
\end{equation}

Now from the definition of $A$ in (I3.13) we find that the ANZC strip for $A$
is $(2+\epsilon,2k-\epsilon)$, but since $A$ is independent of $k$, we can take the
maximal allowed $k$ value, which gives the strip $(2+\epsilon,2N-2-\epsilon)$.
Similarly for $\bar A$ we have $(-2N+2+\epsilon,-2-\epsilon)$.

The defining relation for $Q$, (I3.13), provides an ANZC strip for $Q$ which is
2 units wider in both directions:
\begin{equation}
Q {\rm \ is \ ANZC\  in\ } (\epsilon,2N-\epsilon),
\end{equation}
and analogously
\begin{equation}
\bar Q {\rm \ is \ ANZC\  in\ } (-2N+\epsilon,-\epsilon).
\end{equation}
These strips are consistent with both the fact that $Q$ and $\bar Q$ are
complex conjugates of each other and the crucial relation
\begin{equation}
Q^{[2N]}=\bar Q.
\end{equation}
Therefore, Eq.(I3.16) is still valid if we exclude the real axis from the
domain of definition.

\section{Fourier transformation}

Now the problem with defining the Fourier transform of (the log-derivative of)
$Q$ is that the real line is not in the analyticity strip. But the ${\rm Im}\,
\theta=\pi/2$ line is and there is no problem of defining the Fourier transform
of (the log-derivative of) $Q^+$:
\begin{equation}
\widetilde{Q^+}=q_1.
\end{equation}
Similarly
\begin{equation}
\widetilde{\bar Q^-}=\bar q_1.
\end{equation}
Since
\begin{equation}
Q^{[\alpha]}=(Q^+)^{[\alpha-1]},
\end{equation}
in Fourier space we have
\begin{equation}
\widetilde{Q^{[\alpha]}}=p^{\alpha-1}q_1
\end{equation}
and analogously
\begin{equation}
\widetilde{\bar Q^{[-\beta]}}=p^{1-\beta}\bar q_1.
\end{equation}

Let us now define 
\begin{equation}
\widetilde{Q}=\frac{1}{p}q_1,\qquad{\rm and}\qquad
\widetilde{\bar Q}=p\bar q_1.
\end{equation}
Note that although $\widetilde{Q}$, $\widetilde{\bar Q}$ are not Fourier
transforms of anything, nevertheless we can write the relations
\begin{equation}
\widetilde{Q^{[\alpha]}}=p^\alpha \widetilde{Q},\qquad{\rm and}\qquad
\widetilde{\bar Q^{[-\beta]}}=p^{-\beta}\widetilde{\bar Q}.
\end{equation}
Similarly, instead of the relation $Q^{[2N]}=\bar Q$ , one can take the Fourier transform of 
its equivalent form  
\begin{equation}
Q^{[2N-1]}=\bar Q^-
\end{equation}
since both sides are in their respective analyticity strips to get
\begin{equation}
p^{2N-1}\widetilde{Q}=\frac{1}{p}\widetilde{\bar Q},
\end{equation}
which is of course equivalent to the relation
\begin{equation}
\widetilde{\bar Q}=p^{2N}\widetilde{Q}.
\end{equation}
This relation was used in the derivation of the sausage model NLIE equations in Fourier
space.

We can still apply a procedure of constructing NLIE in Fourier space,
initiated by \cite{Junji1} since (I3.20-21) remain
valid if we interpret them as Fourier space
relations only. However, after eliminating $\tilde Q$ and $\tilde{\bar Q}$,
we arrive at (I3.25-26), where all building blocks are again genuine Fourier
transforms. The results for the sausage model NLIE are thus
unchanged\footnote{The resulting NLIE, eq. (I3.32-3.34), turns out to be in
agreement with the one conjectured by Clare Dunning in \cite{Clare}.}.

\section*{Acknowledgements}
This work was supported by the National Research Foundation of Korea (NRF) grant
(NRF-2016R1D1A1B02007258) (CA), by the Hungarian National Science Fund OTKA (under K116505) (JB), and by Commission IV (Theory) of I.N.F.N. under the grant GAST (FR).
We thank Junji Suzuki for pointing out that one of our steps in the
original derivation was not sufficiently justified. His remark prompted
the present Addendum. We also thank him for valuable discussions during
the preparation of this manuscript.

\appendix

\section{Derivation of analyticity strips}
Using the $Y$-system equations we look for the maximal analyticity strips.
For example $y_1$ can be written as
\begin{equation}
y_1^+=\frac{Y_2}{y_1^-}
\end{equation}
and for $\theta\in(0,1)$ the LHS defines $y_1$ in the strip $(1,2)$. The
numerator on the RHS lives in $(0,1)$ and the denominator in $(-1,0)$. We
already know that this RHS is ANZC so we can conclude that $y_1$ is ANZC also
in $(1,2)$. Similar conclusions can be drawn from the equations
\begin{equation}
y_k^+=\frac{Y_{k-1}Y_{k+1}}{y_k^-} 
\end{equation}
for $k=3,\dots$. However,
we can only conclude that $y_2$ is ANZC in $(1,2-\epsilon)$ from
\begin{equation}
y_2^+=\frac{Y_1Y_3Y_0}{y_2^-}
\end{equation}
because of $Y_0$ in the numerator.
Of course, analogous considerations apply in the negative imaginary direction.

Let us summarize:
\begin{equation}
y_a {\rm \ is \ ANZC\  in\ } (-2,2) \quad{\rm for}\quad a=1,\dots,N\quad a\not=2
;\qquad
y_2 {\rm \ is \ ANZC\  in\ } (-2+\epsilon,2-\epsilon).
\end{equation}
Now continuing this procedure we can convince ourselves that 
\begin{equation}
y_3 {\rm \ is \ ANZC\  in\ } (-3+\epsilon,3-\epsilon),\qquad 
y_4 {\rm \ is \ ANZC\  in\ } (-4+\epsilon,4-\epsilon),
\end{equation}
and so on. In the language of the variables $Z_k$ we have
\begin{equation}
Z_k {\rm \ is \ ANZC\  in\ } (-k+\epsilon,k-\epsilon),\qquad 
k=1,\dots,N-1.
\end{equation}

Finally since the $T$-system functions are defined as the solution of the
basic TBA-like equation
\begin{equation}
T_k^+T_k^-=Z_k,
\end{equation}
they have 1 unit wider strips:
\begin{equation}
T_k {\rm \ is \ ANZC\  in\ } (-k-1+\epsilon,k+1-\epsilon),\qquad 
k=1,\dots,N-1.
\end{equation}

\end{document}